\documentclass[aip,amsmath,amssymb,reprint,
groupedaddress,
]{revtex4-1}
\usepackage[usenames,dvipsnames]{color}
\usepackage{tikz}
\usepackage{graphicx}
\usepackage{dcolumn}
\usepackage{bm}
\usepackage{amsmath}
\usepackage{amsfonts}
\usepackage{psfrag}
\usepackage[bookmarks,colorlinks=false]{hyperref}
\usepackage{mathtools}
\usepackage{accents}

\usepackage[mathcal]{euscript}

\usetikzlibrary{arrows,intersections,decorations.pathmorphing}

\newcommand{\Eq}[1]{Eq.~(\ref{#1})}
\newcommand{\Eqs}[2]{Eqs.~(\ref{#1})-(\ref{#2})}

\begin{document}
	
	\title{Exact solution of polaritonic systems with arbitrary light and matter frequency-dependent losses}

\author{Erika Cortese}
\email{E-mail: e.cortese@soton.ac.uk}
\author{Simone \surname{De Liberato}}
\email{E-mail: s.de-liberato@soton.ac.uk}
\affiliation{School of Physics and Astronomy, University of Southampton, Southampton, SO17 1BJ, United Kingdom}

\begin{abstract}
In this paper we perform the exact diagonalization of a light-matter strongly coupled  system taking into account arbitrary losses via both energy dissipation in the optically active material and photon escape out of the resonator. This allows to naturally treat the cases of couplings with structured reservoirs, which can strongly impact the polaritonic response via frequency-dependent losses or discrete-to-continuum strong coupling.
We discuss the emergent gauge freedom of the resulting theory and provide analytical expressions for all the gauge-invariant observables both in the Power-Zienau-Woolley and the Coulomb representations. 
In order to exemplify the results the theory is finally specialised to two specific cases. In the first one both light and matter resonances are characterised by Lorentzian linewidths, and in the second one a fixed absorption band is also present.
The analytical expressions provided in this paper can be used to predict, fit, and interpret results from polaritonic experiments with arbitrary values of the light-matter coupling and with losses of arbitrary intensity and spectral shape, in both the light and matter channels.

\end{abstract}

	\maketitle
	\section{Introduction}
\begin{table*}
\label{table}
\begin{tabular}{ |c|c|c|c| } 
 \hline
 Symbol & Description & Depending on & First appearance in \\
 \hline\hline
 $k$ & composite wavevector index  & & \Eq{HPZW}\\
 $\omega_k$ & bare photon frequency & $k$ & \Eq{HPZW}\\
 $a_k$ & photon annihilation operator &  $k$ & \Eq{HPZW} \\ 
 $\omega_x$ & bare matter frequency & &\Eq{HPZW}\\
 $b_k$ & matter annihilation operator & $k$ & \Eq{HPZW} \\ 
 $g$ & light-matter coupling strength &  & \Eq{HPZW} \\
 $\tilde{\omega}_x$ & matter frequency renormalised by diamagnetic term & &\Eq{HPZW}\\
 $\omega_{\pm,k}$ & polariton frequencies & $k$ & \Eq{lossless}\\
 $\alpha_k$ & photonic reservoir annihilation operators &  $k$ and $\omega$ & \Eq{alphabeta} \\
 $\beta_k$ & matter reservoir annihilation operators &  $k$ and $\omega$ & \Eq{alphabeta} \\
 $V_k$ & photonic reservoir interaction function &  $k$ and $\omega$ & \Eq{PB} \\
 $Q$ & matter reservoir interaction function & $\omega$ & \Eq{MB}\\
 $\bar{\omega}_k$ & photon frequency renormalised by reservoir & $k$ & \Eq{baromega}\\
 $\bar{\omega}_x$ & matter frequency renormalised by reservoir & $k$ & \Eq{baromega}\\
 $A_k$ & broadened photon operator  & $k$ and $\omega$ & \Eq{A}\\
 $x_k$,$z_k$  & photonic mode Hopfield coefficients & $k$ and $\omega$ &\Eq{A} \\
 $y_k$,$w_k$ & photonic reservoir Hopfield coefficients & $k$, $\omega$ and $\omega' $ &\Eq{A} \\
 $\gamma_k$ & unknown function  & $k$ and $\omega$ & \Eq{yk}\\
 $\chi_k$ & self-energy term for the photonic reservoir   & $k$ and $\omega$ & \Eq{chi}\\
  $B_k$ & broadened matter operator  & $k$ and $\omega$ & \Eq{B}\\
 $x$,$z$  & matter mode Hopfield coefficients & $\omega$ &\Eq{B} \\
 $y$,$w$  & matter reservoir Hopfield coefficients & $\omega$, $\omega'$ &\Eq{B} \\
 $t$ &   self-energy term for the matter reservoir  & $\omega$ & \Eq{t}\\
 $\zeta_k$ & expansion coefficient of the photonic field & $k$ and $\omega$ & \Eq{fieldsab}\\
 $\eta$ & expansion coefficient of the matter field & $\omega$ & \Eq{fieldsab}\\
 $P_{k,j}$ & polariton operator  & $k$, branch $j=\pm$ and $\omega$ & \Eq{P}\\
 $\bar{x}_{k,j}$,$\bar{y}_{k,j}$,$\bar{w}_{k,j}$,$\bar{z}_{k,j}$  & polariton Hopfield coefficients & $k$, branch $j=\pm$ and $\omega$, $\omega' $ &\Eq{P} \\
 $K_{k,j}$ & integral function of the photonic  coefficient  & $k$, branch $j=\pm$ and $\omega$ & \Eq{KJ}\\
 $J_{k,j}$ & integral function of the matter  coefficient  & $k$, branch $j=\pm$ and $\omega$ & \Eq{KJ}\\
 $Z$ & integral function of $ |\eta(\omega)|$ & $k$ and $\omega$ & \Eq{ZZZ}\\
 $W_k$ & integral function of $ |\zeta_k(\omega)|$  & $k$ and $\omega$ & \Eq{ZZZ}\\
 $\mathsf{X}_{j,k}$,$\mathsf{Y}_{j,k}$,$\mathsf{Z}_{j,k}$, $\mathsf{W}_{j,k}$, & Hopfield coefficients of broadened polaritons & $k$, branch $j=\pm$ and $\omega$  & \Eq{XXj} \\
 $\gamma_P$ & photonic reservoir loss rate  & & \Eq{V}\\
 $\omega_P$ & photonic reservoir cut off frequency  & & \Eq{V}\\
 $\gamma_M$ & matter reservoir loss rate  & & \Eq{V}\\
 $\omega_M$ & matter reservoir cut off frequency  & & \Eq{V}\\
 $V^1_k$ & interaction function between light and absorption band &  $k$ and $\omega$ & \Eq{VV} \\
 $\kappa$ & interaction strength to the reservoir & & \Eq{V1}\\
 ${{\bar{\omega}_k}}$ & photonic frequency renormalised by reservoir and continuum & $k$ & \Eq{V1}\\
 $\tilde{\omega}_k$ &  photonic frequency renormalised by continuum  & $k$ & \Eq{tilde}\\
 $F$ & normalised coupling density to the reservoir  & $\omega$ & \Eq{V1}\\
 $\Omega_k$ & effective central frequency for the added reservoir & $k$ and $\omega$ & \Eq{OG}\\
 $\Gamma$ & effective loss rate & $\omega$ & \Eq{OG}\\
  
 \hline
\end{tabular}
\end{table*}

The interaction between discrete energy levels and degrees of freedom with continuum spectra is crucial to the description of any real-world quantum system, in which the coupling with the environment eventually leads to energy and information leakage.
While many powerful perturbative open quantum systems approaches have been developed \cite{breuer_theory_2007}, non-perturbative diagonalization is possible using a procedure due to Fano \cite{fano_effects_1961}. In his landmark paper Fano considered the problem of one discrete level coupled to one continuum.
In the same paper he then moves to consider the cases of multiple discrete levels coupled to one continuum and of one discrete level coupled to multiple continua, showing that both cases can be reduced to the initial one. A short summary of Fano's approach and its generalizations is given in Appendix \ref{appendix:Appz}.

 One important application of the Fano's theory in the many discrete levels-one continuum case is light interacting with a dissipative dielectric, originally developed by Huttner and Barnett (HB) \cite{huttner_quantization_1992}. 
In such a formalism light with a well-defined momentum and polarization propagating in a bulk dielectric is represented in second quantization as an harmonic oscillator. The light is coupled with a discrete optical resonance of the material, itself modeled as an harmonic degree of freedom and coupled to an harmonic reservoir leading to dissipation. 
By diagonalizing the light-matter Hamiltonian one finds two hybrid polaritonic branches, which in the following we will call lower (-) and upper (+) polaritons, coupled to a reservoir through their matter component. As expected, the more matter-like is the polariton the larger losses it will incur, with pure photons very detuned from the material resonance propagating unimpeded in the dielectric. 

Complications arise in systems with boundaries, as traditional cavity quantum electrodynamics (CQED) setups or surface modes. An HB-like diagonalization can still be performed in real space if the photons are supposed to be perfectly trapped in a finite volume \cite{gubbin_real-space_2016}, but in the general case a novel dissipation pathway opens, this time linked with the photonic component of the polaritons: photons can escape out of the system coupling with the free-space photonic continuum.

Such a setup is thus described by two discrete resonances (the photons and the optically active resonance) coupled to two different continua (the material reservoir and the continuum of free space photons). This case was not explicitly treated in the Fano's original paper and, as we will see, it is not possible to trivially apply the method adopted in the other cases. 
Still, various approximate approaches have been developed to deal with open CQED.
Input-output approaches integrate out the system in order to describe relations between the incoming and outgoing fields \cite{savasta_quantum_1996,ciuti_input-output_2006,de_liberato_comment_2014}. Master equations integrate out the environment \cite{beaudoin_dissipation_2011,bamba_recipe_2014}, or at least most of it \cite{iles-smith_environmental_2014}, to describe the internal system dynamics. Some approaches exactly solve the coupling with the propagative electromagnetic field (radiative broadening), while describing phenomenologically matter losses \cite{alpeggiani_quantum_2014,franke_quantization_2019}. It is also possible to use quasinormal mode quantization in order to quantize directly the lossy electromagnetic field \cite{leung_completeness_1994,lalanne_light_2018,franke_quantization_2019}.

This large interest is motivated by the increasing experimental relevance of a rigorous treatment of lossy CQED systems, including the impact of frequency-dependent structured environments. Ever larger values of the light-matter interaction energy \cite{ballarini_polaritonics_2019} have in-fact allowed us to access non-perturbative coupling regimes as
the ultrastrong  \cite{forn-diaz_ultrastrong_2019,frisk_kockum_ultrastrong_2019,anappara_signatures_2009,scalari_ultrastrong_2013} or the very strong ones \cite{khurgin_excitonic_2001,brodbeck_experimental_2017}. In these regimes the polaritonic modal shifts are comparable to other energy scales, and frequency-independent approximation can dramatically fail.
In particular polaritonic discrete resonances can interact with continua, with multiple theoretical \cite{averkiev_light-matter_2007,citrin_microcavity_2003,cortese_strong_2019,parish_excitons_2021,cao_strong_2021} and experimental \cite{liu_quantum_2017,mueller_deep_2020,cortese_excitons_2021,rajabali_polaritonic_2021} efforts having studied the possibility of strong coupling taking the continuum into account.

An analytical solutions extending the Fano's approach would be useful in this context, in order to be able to study the quantum properties of systems in the presence of generic couplings, environments, and loss channels.
It would allow for quantitative modeling of the lineshape of plasmonic systems once the loss channels in the metals are known \cite{khurgin_how_2015}.
Such an approach was derived in Ref. \cite{de_liberato_virtual_2017} to calculate the quantum properties of the ground state at arbitrary values of the system-reservoir coupling.
In this work it was shown how an unphysical degree of freedom appears in the theory due to the presence of two coupling continua, and how a solution can be obtained through an arbitrary gauge fixing.
The method has also been more recently used in Ref. \cite{rajabali_polaritonic_2021} to reproduce experimental data in which polaritonic nonlocality created a broad absorption band above the bare photonic frequency.

Our aim in this paper is to improve over such a contribution and develop a full, general, and usable analytic theory for polaritonic systems with arbitrary couplings to the environment. Such an improvement will take four different forms. The first and non-negligible one is that the theory will be clearly laid down in the paper, while in Refs. \cite{de_liberato_virtual_2017,rajabali_polaritonic_2021} the derivation is at most sketched.
The second, more important reason is that the pure gauge nature of the extra degree of freedom was not proven but only assumed.
The third is that the theory was developed in the Coulomb representation, which has since then be shown to be not always correct at arbitrarily large coupling strengths \cite{de_bernardis_breakdown_2018,stokes_gauge_2019,di_stefano_resolution_2019} for systems in which only few material resonances are considered.  
Finally, the theory previously used, although applicable to model reservoirs of arbitrary spectral shapes, requires a renormalization procedure. It is thus not directly applicable to cases beyond the presence of a simple Lorentzian broadening.

We will implement these improvements by developing explicitly the theory in the Power-Zienau-Woolley (PZW) representation. Calculating results for the gauge-invariant physical observables without fixing the extra degrees of freedom we will show their fully gauge nature. Our results will thus provide a proof of the  approach from Ref. \cite{de_liberato_virtual_2017}, while demonstrating gauge fixing is not necessary.
Finally we will provide a recipe to add arbitrary frequency-dependent reservoirs.
Given this work's technical nature we describe the calculations in details. Equivalent results for the Coulomb representation are reported for completeness in the Appendix \ref{appendix:AppA}. All the mathematical symbols used are listed in Table \ref{table}.

The paper is organised as follows.
In Sec. \ref{dissipationless} we introduce and diagonalise the dissipationless polaritonic Hamiltonian. This will be useful to introduce the problem and to extract the discrete polaritonic dispersion which we will then use to interpret the results of the dissipative theory.
Note that in this paper we will always start from Hamiltonians. Their derivation from Lagrangians can be found in Ref. \cite{gubbin_real-space_2016} for the PZW case and in Ref. \cite{huttner_quantization_1992} for the Coulomb case.
In Sec. \ref{reservoirs} we introduce the light and matter reservoirs and we diagonalise independently the light and matter sectors of the full Hamiltonian, into a broadened matter resonance interacting with a broadened photonic one.
In Sec. \ref{fullH} we solve the full Hamiltonian, describing the problem with gauge ambiguities, and derive expressions for the gauge-invariant observables.
In Sec. \ref{Lorentz} we specialise the model to the case of Lorentzian resonances. 
In Sec. \ref{Intrinsec} we provide the recipe to add arbitrary reservoirs to the Lorentzian broadening, and present the results in the case of a fixed absorption band.
	
	\section{Diagonalization in the lossless case}	
	\label{dissipationless}
	
	We start here by introducing and diagonalizing the lossless Hamiltonian for a photon field of dispersion $\omega_k$ indexed by the composite wavevector index $k$, which incorporates also polarization and all other relevant conserved quantum numbers.
	Such a field is described by the bosonic annihilation operator $a_k$ coupled to a matter excitation, of fixed frequency $\omega_x$, described by the bosonic annihilation operator $b_k$. We neglect here nonlocality due to the dispersion of the material resonance \cite{ciraci_probing_2012,gubbin_optical_2020,rajabali_polaritonic_2021}.
	
	The PZW light-matter Hamiltonian for the system described above is
	\begin{eqnarray}
	\label{HPZW}
		H_{\text{LM}}&=& \sum_k \left (\omega_k a^{\dagger}_k a_k+\omega_x \, b^{\dagger}_k b_k \right )
		 	+
		\sum_k \frac{ g^2}{\omega_x}
		\left[b^{\dagger}_k+b_k \right]^2 \nonumber\\&&+i \sum_k g \sqrt{\frac{\omega_k}{\omega_x}}\left[a^{\dagger}_k-a_k \right]
		\left[b^{\dagger}_k+b_k\right],
	 \label{aa}
	\end{eqnarray}
	with $g$ the resonant light-matter coupling strength and the operators obeying bosonic commutator relations
	\begin{align}
	\left[a_k,a_{k'} \right] =\left[b_k,b_{k'} \right]   =\delta_{k,k'},
	\end{align}
with $\delta_{k,k'}$ a Kronecker delta.	
The second term in \Eq{HPZW} is the square $P^2$ term which can be removed by performing a Bogoliubov rotation of the matter component of the Hamiltonian, renormalizing the bare matter frequency, as
\begin{eqnarray}
\tilde{\omega}_x^2=\omega_x^2+4g^2.
\end{eqnarray}
The Hamiltonian in \Eq{HPZW} then takes the simpler form 
	\begin{eqnarray}
	\label{HPZWS}
		H_{\text{LM}}&=& \sum_k \left (\omega_k a^{\dagger}_k a+\tilde{\omega}_x \,b^{\dagger}_k b_k \right)
		+\\&& i \sum_k g \sqrt{\frac{\omega_k}{\tilde{\omega}_x}} \left[a^{\dagger}_k-a_k \right]
		\left[ b_{k}^{\dagger}+ b_k\right]. \nonumber
	\end{eqnarray}
The secular equation of such an Hamiltonian reads
	\begin{equation} \label{dis}
\omega^2-\omega_k^2= \frac{4 g^2\omega_k^2}{\omega^2-\tilde{\omega}_x^2},
\end{equation}	
leading to the frequencies of the two polariton branches ($\pm$) for each value of the wavevector $k$ 
\begin{equation}
\label{lossless}
\omega_{\pm,k}= \frac{1}{\sqrt{2}}\sqrt{\omega_k^2+ \tilde{\omega}_x^2 \pm \sqrt{(\omega_k^2-\tilde{\omega}_x^2)^2 + 16 g^2\omega_k^2}}.
\end{equation}

\section{Diagonalization  of photonic and matter reservoirs}	
\label{reservoirs}

We now pass to introduce the dissipation in the picture by defining two reservoirs in which photons and matter excitations can be lost. Those reservoirs, which model respectively the continuum of free-space photons and the continuum of phononic and electronic degrees of freedom in the material are modeled
 as ensambles of harmonic oscillators indexed by the continuum frequency $\omega$. Their annihilation operators are $\alpha_k(\omega)$ and $\beta_k(\omega)$ respectively, with bosonic commutator relations
 \begin{eqnarray} \label{alphabeta}
 &&\left[\alpha_k(\omega),\alpha_{k'}^{\dagger}(\omega') \right]\!=\!
 \left[\beta_k(\omega),\beta_{k'}^{\dagger}(\omega') \right]\!=
 \delta_{k,k'}\delta(\omega\!-\!\omega').\quad
 \end{eqnarray}
 The total Hamiltonian now takes the form
 \begin{eqnarray}
 H_\text{T}&=&H_{g}+H_{\text{PB}}+H_{\text{MB}},
 \end{eqnarray}
with the Hamiltonians for the photonic and matter sectors and their interaction in the form
\begin{align}
     \label{PB}
 H_{\text{PB}}&= \sum_{k} \left [ \bar{\omega}_k a_k^\dagger a_k +\int\!d\omega \omega \alpha_k^\dagger(\omega) \alpha_k(\omega) \right. +\\& \left.\frac{1}{2} \sum_{k}  \int\!d\omega (a_k^\dagger + a_k) \left (V_k(\omega) \alpha_k^\dagger(\omega)+ V_k^*(\omega)\alpha_k(\omega) \right) \right],\nonumber\\ \label{MB}
H_{\text{MB}} &= \sum_{k} \left [ \bar{\omega}_x b_k^\dagger b_k+ \int\!d\omega \omega \beta_k^\dagger(\omega) \beta_k(\omega) \right. +\\& \left. \frac{1}{2}  \int\!d\omega (b_k^\dagger + b_k) \left (Q(\omega') \beta_k^\dagger(\omega')+ Q^*(\omega')\beta(\omega') \right) \right], \nonumber \\
\label{Hg}
H_{g}&=i \sum_k g \sqrt{\frac{\bar{\omega}_k}{\bar{\omega}_x}}\left[a^{\dagger}_k-a_k \right]
		\left[b^{\dagger}_k+b_k\right],
\end{align}
where $V_k(\omega)$ and $Q(\omega)$ are the interaction functions modeling the interaction of, respectively, the photonic mode and the matter excitation with their respective reservoirs, and   the bare light and matter resonances are dressed by the coupling as
\begin{eqnarray} \label{baromega}
\bar{\omega}_k^2&=& \omega_k^2+ \int_0^\infty d\omega \frac{|V_k(\omega)|^2\bar{\omega}_k}{\omega},\nonumber\\
\bar{\omega}_x^2&=& \tilde{\omega}_x^2+ \int_0^\infty d\omega \frac{|Q(\omega)|^2 \bar{\omega}_x}{\omega}.
\end{eqnarray}

We can diagonalise the photonic Hamiltonian $H_{\text{PB}}$ in \Eq{PB} introducing the  bosonic operators describing broadened photons $A_k(\omega) $, 
\begin{eqnarray} \label{A}
A_k(\omega)&=& x_k(\omega) a_k+ z_k(\omega) a_k^\dagger + \\&&\int\!d\omega' \left [ y_k(\omega,\omega') \alpha_k(\omega')+ w_k(\omega,\omega') \alpha_k^\dagger (\omega') \right ], \nonumber 
\end{eqnarray}
whose coefficients can be found via HB diagonalization, illustrated in more details in Appendix \ref{appendix:AppF}.
From the eigenequation 
\begin{eqnarray}
\omega A_k(\omega)=\left[A_k(\omega),H_{\textrm{PB}}\right],
\end{eqnarray}
the resulting system reads
\begin{eqnarray}
x_k(\omega) \left(\omega\! -\! \bar{\omega}_k \right )&=&\frac{1}{2}\int_0^\infty d\omega' \left [ y_k(\omega,\omega') V_k(\omega')- \right. \nonumber\\ && \left.w_k(\omega,\omega') V_k^*(\omega') \right],\label{xxk} \\
z_k(\omega) \left(\omega\! +\! \bar{\omega}_k\right )&=&\frac{1}{2}\int_0^\infty d\omega' \left [ y_k(\omega,\omega') V_k(\omega')- \right. \nonumber\\ && \left. w_k(\omega,\omega') V_k^*(\omega') \right],\\
y_k(\omega,\omega') \left(\omega\!-\!\omega'\right)&=&\frac{1}{2}\left [ x_k(\omega) - z_k(\omega)\right] V_k^*(\omega'), \label{Yyk}\\
w_k(\omega,\omega') \left(\omega\!+\!\omega'\right)&=&\frac{1}{2}\left [ x_k(\omega) - z_k(\omega)\right] V_k(\omega').
\label{wwk}
\end{eqnarray}
Such a system cannot be trivially solved eliminating one-by-one its unknowns because, under the hypotesis that the eigenfrequency $\omega$ falls into the photonic reservoir continuum, there will always be a value of $\omega'=\omega$ which makes the left-hand-side of  and \Eq{Yyk} vanish. This is in stark contrast with the discrete case in which coupled modes are never degenerate with bare resonances \cite{de_liberato_light-matter_2014,todorov_dipolar_2015}.
The system can nevertheless be solved in the distribution sense as
\begin{eqnarray} \label{yk}
y_k(\omega,\omega')&=&\left [ P \left( \frac{1}{\omega\!-\!\omega'}\right)+ \gamma_k(\omega) \delta(\omega\!-\!\omega') \right] \times \\
\nonumber
&&\frac{1}{2}\left [ x_k(\omega) - z_k(\omega)\right] V_k^*(\omega'),
\end{eqnarray}
where $P$ is the principal value and $\gamma_k(\omega)$ is an unknown function, which can be fixed by imposing the bosonic commutation relation
\begin{eqnarray}
\left[A_k(\omega),A_{k'}(\omega')^{\dagger} \right]&=&\delta_{k,k'}\delta(\omega\!-\!\omega').
\end{eqnarray}
After some manipulations we can solve the system in \Eqs{xxk}{wwk} arriving to the  following expressions for the coefficients
\begin{eqnarray} \label{x}
 x_k(\omega)&=&\frac{  \omega+\bar{\omega}_k }{2}  \frac{V_k(\omega)}{\omega^2-\bar{\omega}_k^2 \chi_k(\omega)},\\ \label{z}
 z_k(\omega)&=&\frac{  \omega-\bar{\omega}_k }{2}  \frac{V_k(\omega)}{\omega^2-\bar{\omega}_k^2 \chi_k(\omega)},\nonumber\\
 y_k(\omega,\omega')&=& \delta(\omega\!-\!\omega')+\nonumber\\&& \frac{\bar{\omega}_k}{2} \frac{V_k(\omega')}{\omega\!-\!\omega'-i 0^+} \frac{V_k(\omega)}{\omega^2-\bar{\omega}_k^2 \chi_k(\omega)},\nonumber\\
 w_k(\omega,\omega')&=&  \frac{\bar{\omega}_k}{2}  \frac{V_k(\omega')}{\omega\!+\!\omega'} \frac{V_k(\omega)}{\omega^2-\bar{\omega}_k^2 \chi_k(\omega)},\nonumber 
\end{eqnarray}
with 
\begin{eqnarray} \label{chi}
\chi_k(\omega)=1-\frac{1}{2 \bar{\omega}_k} \int_{-\infty}^\infty d\omega' \frac{\mathcal{V}_k(\omega')}{\omega'-\omega+i 0^+},
\end{eqnarray}
and we defined $\mathcal{V}_k(\omega)$  the odd analytic extension of $|V_k(\omega)|^2$ in the negative frequency range. 

Exactly the same procedure can be applied to the Hamiltonian in \Eq{MB} describing the matter sector $H_{\text{MB}}$, by introducing the bosonic operator for the broadened optically active resonance
\begin{eqnarray} \label{B}
B_k(\omega)&=& \bar{x}(\omega) b_k+ \bar{z}(\omega) b_k^\dagger + \\&&\int\!d\omega' \left [ \bar{y}(\omega,\omega') \beta_k(\omega')+ \bar{w}(\omega,\omega') \beta_k^\dagger (\omega') \right ]. \nonumber
\end{eqnarray}
The solution is in the analogous form
\begin{eqnarray} \label{ovx}
 \bar{x}(\omega)&=&\frac{  \omega+\bar{\omega}_x }{2}  \frac{Q(\omega)}{\omega^2-\bar{\omega}_x^2 t(\omega)},\\ \label{ovz}
 \bar{z}(\omega)&=&\frac{  \omega-\bar{\omega}_x }{2}  \frac{Q(\omega)}{\omega^2-\bar{\omega}_x^2 t(\omega)},\nonumber\\
 \bar{y}(\omega,\omega')&=& \delta(\omega\!-\!\omega')+\nonumber\\&&\nonumber  \frac{\bar{\omega}_x}{2} \frac{Q(\omega')}{\omega\!-\!\omega'-i 0^+} \frac{Q(\omega)}{\omega^2-\bar{\omega}_x^2 t(\omega)},\nonumber\\
 \bar{w}(\omega,\omega')&=&  \frac{\bar{\omega}_x}{2}  \frac{Q(\omega')}{\omega\!+\!\omega'}  \frac{Q(\omega)}{\omega^2-\bar{\omega}_x^2 t(\omega)},\nonumber
\end{eqnarray}
with 
\begin{eqnarray} \label{t}
t(\omega)=1-\frac{1}{2
\bar{\omega}_x} \int_{-\infty}^\infty d\omega' \frac{\mathcal{Q}(\omega')}{\omega'-\omega+i 0^+}.
\end{eqnarray}
As for its photon counterpart, we defined $\mathcal{Q}(\omega)$  the odd analytic extension of $|Q(\omega)|^2$ in the negative frequency range. 

We can now recover the inverse transformations for the bare operators in term of the broadened ones, 
\begin{eqnarray}
a_k&=& \int_0^\infty d\omega' \left [x_k^*(\omega') A_k(\omega')  - z_k(\omega') A_k^\dagger(\omega') \right],\\
a_k^\dagger&=& \int_0^\infty d\omega' \left [x_k(\omega') A_k^\dagger(\omega')  - z_k^*(\omega') A_k(\omega') \right],\nonumber\\
b_k&=&\int_0^\infty d\omega' \left[ \bar{x}^*(\omega) B_k(\omega')-\bar{z}(\omega) B_k^\dagger(\omega') \right],\nonumber\\
b^\dagger_k&=&\int_0^\infty d\omega' \left[ \bar{x}(\omega) B_k^\dagger (\omega')-\bar{z}^*(\omega) B_k(\omega') \right],\nonumber
\end{eqnarray}
and write the bare field operators as superpositions of the broadened fields 
\begin{eqnarray}
\label{fieldsab}
i\left (a_k\!-\!a_k^\dagger \right)&=&\frac{1}{\sqrt{\bar{\omega}_k}} \int_0^\infty d\omega \left [ \zeta_k(\omega) A_k^\dagger(\omega) + \zeta_k^*(\omega) A_k(\omega) \right], \nonumber\\ 
 \left(b^\dagger_k\!+\! b_k \right)&=&\sqrt{\bar{\omega}_x}\int_0^\infty d\omega \left [\eta(\omega) B_k^\dagger(\omega)+\eta^*(\omega) B_k(\omega) \right ],\nonumber \\ 
\end{eqnarray}
where the expansion coefficients of the bare fields upon the broadened operators are given by the expressions
\begin{eqnarray} \label{ze}
\zeta_k(\omega)&=&-i \sqrt{\bar{\omega}_k} \left[x_k(\omega)+z_k(\omega)\right]=-i\frac{V_k(\omega) \omega \sqrt{\bar{\omega}_k }}{\omega^2-\bar{\omega}_k^2 \chi_k(\omega)},\nonumber \\ \label{et}
\eta(\omega)&=& \frac{1} {\sqrt{\bar{\omega}_x}} \left [\bar{x}(\omega)-\bar{z}(\omega) \right]=  \frac{Q(\omega) \sqrt{\bar{\omega}_x }}{\omega^2-\bar{\omega}_x^2t(\omega)}.
\end{eqnarray}

\section{Diagonalization of the full light-matter Hamiltonian}
\label{fullH}

After substituting the field operators in \Eq{fieldsab} into the coupling Hamiltonian from \Eq{Hg}, the full light-matter Hamiltonian can be written in the form 
\begin{align}  \label{HTOT}
H_\text{T}&=  \sum_{k} \int_0^\infty d\omega\omega A_k^\dagger(\omega)A_k(\omega) + \int_0^\infty d\omega \omega B_k^\dagger(\omega) B_k(\omega) \nonumber \\ 
&+ \sum_{k} g \int_0^\infty d\omega  \int_0^\infty d\omega'  \left [ \zeta_k(\omega) A_k^\dagger(\omega) + \zeta_k^*(\omega) A_k(\omega)  \right] \nonumber \\& \times \left [ \eta(\omega') B_k^\dagger(\omega')+ \eta^*(\omega') B_k(\omega') \right], 
\end{align}
which describes the  broadened photonic mode coupled to the broadened material resonance.
Similarly to what done previously, the Hamiltonian can be diagonalised by introducing the operators for two polariton branches $P_{j}(\omega)$ with $j=\pm$
\begin{eqnarray} \label{P}
P_{k,j}(\omega)&=&\int_0^\infty d\omega' \left [\tilde{x}_{k,j}(\omega,\omega') A_k(\omega')+ \tilde{z}_{k,j}(\omega,\omega') A_k^\dagger(\omega') + \right.\nonumber\\&& 
\left.\tilde{y}_{k,j}(\omega,\omega') B_k(\omega') +\tilde{w}_{k,j}(\omega,\omega') B_k^\dagger(\omega') \right],
\end{eqnarray}
defined as arbitrary linear superpositions of the bare modes at all frequencies.
By exploiting the eigenequation 
\begin{eqnarray}
\omega P_{k,j}(\omega)=\left[ P_{k,j}(\omega), H_\text{T} \right ],
\end{eqnarray}
we arrive to a system of equations analogous to \Eqs{xxk}{wwk}
\begin{eqnarray}
&&\tilde{x}_{k,j}(\omega,\omega')(\omega\!-\!\omega' )= g \, \zeta_k^*(\omega') \int\!d\omega''  \frac{2\omega''\eta(\omega'')}{\omega\!+\!\omega''}  \tilde{y}_{k,j}(\omega,\omega''),\nonumber \\
&& \tilde{y}_{k,j}(\omega,\omega')\left (\omega\!-\!\omega' \right)= g \,\eta^*(\omega')  \int\!d\omega'' \frac{2\omega''\zeta_k(\omega'')}{\omega\!+\!\omega''}  \tilde{x}_{k,j}(\omega, \omega'') ,\nonumber \\
&& \tilde{x}_{k,j}(\omega,\omega') \left (\omega\!-\!\omega' \right)\zeta_k(\omega')= \tilde{z}_{k,j}(\omega,\omega')\left (\omega\!+\!\omega' \right) \zeta_k^*(\omega'),\nonumber \\
&& \tilde{y}_{k,j}(\omega,\omega') \left (\omega\!-\!\omega' \right)\eta(\omega')=  \tilde{w}_{k,j}(\omega,\omega')\left (\omega\!+\!\omega' \right) \eta^*(\omega').
\label{xyzw}
\end{eqnarray}
In order to put the system in \Eq{xyzw} in a form apt to be manipulated and solved
we introduce two unknown integral functions of the diagonalization coefficients, which as we will see play the role of photonic and material amplitudes of the polaritonic field
\begin{eqnarray} \label{KJ}
K_{k,j}(\omega)&=&\int\!d\omega' \frac{2\omega'}{\omega + \omega'} \zeta_k(\omega') \tilde{x}_{k,j}(\omega,\omega'), \\
J_{k,j}(\omega)&=&\int\!d\omega' \frac{2 \omega'}{\omega\!+\!\omega'} \eta(\omega') \tilde{y}_{k,j}(\omega,\omega'),\nonumber
\end{eqnarray}
and two known integral functions of the coupling coefficients
\begin{eqnarray} \label{ZZZ}
W_k(\omega)&=&P\int\!d\omega'  \frac{2\omega'}{\omega^2-\omega'^2} |\zeta_k(\omega')|^2,\\
Z(\omega)&=&P\int\!d\omega'  \frac{2\omega'}{\omega^2-\omega'^2} |\eta(\omega')|^2.\\
\nonumber
\end{eqnarray}

Note that, notwithstanding the apparent symmetry, the functions related to the photonic component $|K_{k,j}(\omega)|^2$ and $W_k(\omega)$, have different units from those of the matter part $|J_{k,j}(\omega)|^2$ and $Z(\omega)$. The former are pure numbers, while the latter are times squared. This is due to the specific dependence of the light and matter fields upon their frequency in the PZW representation, clearly visible in \Eq{fieldsab}.

We then solve \Eq{xyzw} for the unknown coefficients
\begin{eqnarray} \nonumber
\tilde{y}_{k,j}(\omega,\omega') &=& \left [ P \left (\frac{1}{\omega\!-\!\omega'} \right) + s_{y,k,j} (\omega) \delta (\omega\!-\!\omega')\right]\! \times  \\
&& g \, \eta^*(\omega')  K_{k,j}(\omega), \nonumber \\ 
\tilde{x}_{k,j}(\omega,\omega') &=& \left [ P \left (\frac{1}{\omega\!-\!\omega'} \right) + s_{x,k,j} (\omega) \delta (\omega\!-\!\omega')\right ]  \times \nonumber \\&& g \, \zeta_k^*(\omega') J_{k,j}(\omega) ,\nonumber
 \nonumber \\
\tilde{z}_{k,j}(\omega,\omega') &=& \frac{1}{\omega\!+\!\omega'} g \, \zeta_k(\omega') J_{k,j}(\omega) ,\nonumber\\
 \tilde{w}_{k,j}(\omega,\omega')&=&\frac{1}{\omega\!+\!\omega'} g \, \eta(\omega')  K_{k,j}(\omega). \label{xJ}
\end{eqnarray}
The crucial difference between this system of equations and the one obtained in the single-continuum case in \Eqs{xxk}{wwk} is that here both bare modes have continuum spectra and thus both the first and the second equations in \Eq{xyzw} diverge.
We have therefore to introduce two unknown functions $s_{x,k,j}(\omega)$ and $s_{y,k,j}(\omega)$ for each value of frequency, wavevector, and for each polaritonic branch. From \Eq{xJ} we can see that those functions allow us to arbitrarily fix the equal-frequency mixing between coupled and uncoupled modes.
 We are thus led to add four different functions at fixed wavevector and frequency, but the commutator relations
\begin{eqnarray}
\label{Pcomm}
\left[ P_{k,j}(\omega),P_{k',j'}^{\dagger}(\omega') \right]&=&\delta_{k,k'}\delta_{j,j'}\delta(\omega\!-\!\omega'),
\end{eqnarray}
represent only three new equations, 
for $j,j'=-$ and  $j,j'=+$ (normalization) and  $j=-$, $j'=+$ (orthogonality). This
leaves a free function corresponding to a $k$- and $\omega$-dependent rotation in the space of the two degenerate polaritonic modes. More generally in the presence of $L$ continua, we would add $L^2$ unknown functions, and obtain $L$ normalization conditions and $\frac{L(L-1)}{2}$ orthogonality conditions, leaving $\frac{L(L-1)}{2}$ quantities to be determined, which is the dimension of the $O(L)$ group. An element of such a group corresponds to a rigid rotation in the space of the $L$ $P_{k,j}(\omega)$ modes at fixed $k$ and $\omega$, which would leave \Eq{Pcomm} unchanged.

We can also express the polariton operators in terms of the bare modes  inserting \Eq{A} and \Eq{B} into \Eq{P}
\begin{eqnarray}
P_{j,k}(\omega)&=& \mathsf{X}_{j,k}(\omega) a_k + \mathsf{Z}_{j,k}(\omega) a_k^\dagger \!+\! \\
&&\mathsf{Y}_{j,k}(\omega) b_k \!+ \!\mathsf{W}_{j,k}(\omega) b_k^\dagger +  \nonumber\\
&&\int_0^\infty\! d\omega' \left [ \mathcal{X}_{j,k}(\omega,\omega') \alpha_k(\omega')\! +\! \mathcal{Z}_{j,k}(\omega,\omega') \alpha_k^\dagger(\omega') \right] + \nonumber\\
&&\int_0^\infty \! d\omega' \left [ \mathcal{Y}_{j,k}(\omega,\omega') \beta_k(\omega')\! +\! \mathcal{W}_{j,k}(\omega,\omega') \beta_k^\dagger(\omega') \right] \nonumber , 
\end{eqnarray}
with the most relevant coefficients having the form
\begin{eqnarray}\label{XXj}
\mathsf{X}_{k,j}(\omega)&=&\int_0^\infty d\omega' \left [ \tilde{x}_{k,j}(\omega,\omega') x_k(\omega') + \tilde{z}_{k,j}(\omega,\omega') z_k^*(\omega')  \right], \nonumber \\  
\mathsf{Z}_{k,j}(\omega)&=&\int_0^\infty d\omega' \left [ \tilde{x}_{k,j}(\omega,\omega') z_k(\omega') + \tilde{z}_{k,j}(\omega,\omega') x_k^*(\omega')  \right],\nonumber\\ 
\mathsf{Y}_{k,j}(\omega)&=&\int_0^\infty d\omega' \left [ \tilde{y}_{k,j}(\omega,\omega') \bar{x}(\omega') + \tilde{w}_{k,j}(\omega,\omega') \bar{z}^*(\omega')  \right], \nonumber  \\
\mathsf{W}_{k,j}(\omega)&=&\int_0^\infty d\omega' \left [ \tilde{y}_{k,j}(\omega,\omega') \bar{z}(\omega') + \tilde{w}_{k,j}(\omega,\omega') \bar{x}^*(\omega')  \right].\nonumber \\
\end{eqnarray}
Writing explicitly the coefficients as in Eqs.(\ref{XXj})  we can at this point find the relations linking the light and matter component of the polaritons to the relevant Hopfield coefficients
\begin{eqnarray} \label{Mco}
K_{k,j}(\omega)&=&  -i\sqrt{\bar{\omega}_k}\left[\mathsf{X}_{k,j}(\omega)+\mathsf{Z}_{k,j}(\omega) \right],\\
J_{k,j}(\omega)&=&\frac{1}{\sqrt{\bar{\omega}_x}}\left[
 \mathsf{Y}_{k,j}(\omega)-\mathsf{W}_{k,j}(\omega)\right]. \nonumber
\end{eqnarray}
The bare photonic and matter field operators can then be expressed in terms of the broadened polaritons
\begin{eqnarray}
i (a_k- a_k^\dagger)&=&
\frac{1}{\sqrt{\bar{\omega}_k}}\! \int^{\infty}_0 \! d\omega \!  \sum_{j=\pm} \left[ K^*_{k,j}(\omega)  P_{k,j}(\omega) + \right.\\
&&\left.
K_{k,j}(\omega)  P_{k,j}^\dagger(\omega) \right ], \nonumber \\
(b_k+b_k^\dagger)&=& \sqrt{\bar{\omega}_x}\int^{\infty}_0  d\omega  \sum_{j=\pm}    \left[ J^*_{k,j}(\omega) P_{k,j}(\omega)+\right.\\ &&\left. J_{k,j} (\omega) P_{k,j}^\dagger(\omega) \right] \nonumber,
\end{eqnarray}
recognising in the expressions in \Eq{KJ} the photonic and matter components of the broadened polaritonic modes.
By expressing the coefficient $\tilde{y}_{k,j}(\omega,\omega')$ as in \Eq{xJ}, we also arrive to a relation between the three quantities defined in \Eq{KJ}
\begin{eqnarray} 
J_{k,j}(\omega)=K_{k,j}(\omega) g\left[ Z(\omega) + s_{y,k,j}(\omega) |\eta(\omega)|^2 \right].
\end{eqnarray}
Inserting the second of \Eq{xJ} into the first of \Eq{KJ}, we derive the two relations ($j=\pm$) 
\begin{eqnarray} \label{1d}
 g^2\!\left[ Z(\omega)\!+\! s_{y,k,j}(\omega) |\eta(\omega)|^2 \right]\left[ W_k(\omega) \!+\! s_{x,k,j}(\omega) |\zeta_k(\omega)|^2 \right]\!=\!1, \nonumber\\
\end{eqnarray}
while from \Eq{Pcomm} we derive the three equations
\begin{eqnarray} \label{d}
&& g^2 \left \{|\eta(\omega)|^2 \left [\pi^2+ s_{y,k,j}(\omega) s^*_{y,k,j'}(\omega) \right] J_{k,j}(\omega) J^*_{k,j'}(\omega)+\right. \\  
&& |\zeta_k(\omega)|^2  \left.\left [\pi^2+ s_{x,k,j}(\omega) s^*_{x,k,j'}(\omega) \right] K_{k,j}(\omega) K^*_{k,j'}(\omega)\right \}= \delta_{j,j'},\nonumber
\end{eqnarray}
 for $j,j'=-$, $j,j'=+$ and  $j=-,j'=+$ respectively.
We are thus left, as anticipated, with an underdetermined set of 5 equations (2 from \Eq{1d} and 3 from \Eq{d}) in six unknowns: the  $K_{k,j}(\omega)$ and the 2 functions $s_{x,k,j}(\omega)$, $s_{y,k,j}(\omega)$ for each of the two values of $j$.
 
In Ref. \cite{de_liberato_virtual_2017} this problem was solved arbitrarily fixing $s_{x,k,-}(\omega)=0$ and then solving the remaining equations. 
Here instead we adopt a different approach, solving directly \Eqs{1d}{d} for the gauge-invariant quantities.
 Due to the arbitrariness of the gauge choice, it is in-fact meaningless to distinguish between lower and upper polariton operators as only the gauge invariant quantities are the total photonic and material polaritonic components
\begin{eqnarray}
\label{ginv}
|K_k(\omega)|^2&=\sum_{j=\pm}|K_{k,j}(\omega)|^2,\\
|J_k(\omega)|^2&=\sum_{j=\pm}|J_{k,j}(\omega)|^2.\nonumber
\end{eqnarray}

Although the solution is algebraically cumbersome, it is possible to find the analytic expressions  for the total light and matter polaritonic components, which are the central result of this paper
\begin{widetext}
\begin{eqnarray}\label{final}
|K_k(\omega)|^2&=&\!\frac{
g^2\left[W_k^2(\omega)+\pi^2|\zeta_k(\omega)|^4\right]|\eta(\omega)|^2+|\zeta_k(\omega)|^2}{\left[g^2W_k(\omega)Z(\omega)-1\! \right]^2+\!g^4\pi^2\left[|\eta(\omega)|^4W_k^2(\omega)
\!+\!|\zeta_k(\omega)|^4Z^2(\omega) \right]\!
+\!g^2\pi^2|\zeta_k(\omega)|^2
|\eta(\omega)|^2\left[2+g^2\pi^2|\zeta_k(\omega)|^2
|\eta(\omega)|^2 \right]
},\quad \\
|J_k(\omega)|^2&=&\frac{ g^2
\left[Z^2(\omega)+\pi^2|\eta(\omega)|^4\right]|\zeta_k(\omega)|^2+
|\eta(\omega)|^2
}
{\left[g^2W_k(\omega)Z(\omega)-1\! \right]^2+g^4\pi^2\left[|\eta(\omega)|^4W_k^2(\omega)
\!+|\zeta_k(\omega)|^4Z^2(\omega) \right]\!
+g^2\pi^2|\zeta_k(\omega)|^2
|\eta(\omega)|^2\left[2+g^2\pi^2|\zeta_k(\omega)|^2
|\eta(\omega)|^2 \right]
}.\quad \nonumber
\end{eqnarray}
\end{widetext}
 
Note that the gauge extra variable disappears as expected from \Eq{final}, thus proving its pure gauge nature and as a consequence the  correctness of the results from Ref. \cite{de_liberato_virtual_2017}, where those same quantities had been calculated by an arbitrary gauge fixing.

\section{Lorentzian resonances}
\label{Lorentz}
In order to analytically evaluate the functions introduced above, we need to specify a model for the coupling between the bare modes and the photonic and matter reservoirs. 
We start by writing a model for homogeneously broadened resonances, thus reproducing the optical response of a Lorentz dielectric model. There have been claims that such a dielectric model cannot be modeled in the HB theory \cite{dutra_permittivity_1998}. This is nevertheless false as we will now demonstrate constructively. The problem in Ref. \cite{dutra_permittivity_1998} is that the authors do not recognise the need to introduce a divergent renormalized frequency.
We will consider couplings of the form
\begin{eqnarray} \label{V}
|V_k(\omega)|^2&=&\frac{\omega \bar{\omega}_k}{q_P+\omega_{P}}\Theta(\omega_P-\omega),\\\nonumber
|Q(\omega)|^2&=&\frac{\omega \bar{\omega}_x}{q_M+\omega_M}\Theta(\omega_M-\omega),
\end{eqnarray}
with $q_P=\pi \omega_k^2/2 \gamma_P$ and $q_M=\pi \tilde{\omega}_x^2/2 \gamma_M$, where we have introduced cut off frequencies  $\omega_{P}$ and $\omega_M$, the photonic and matter loss rates, $\gamma_{P}$ and $\gamma_{M}$ and $\Theta$ is the Heaviside function.
In \Eq{V} also appear the renormalised frequencies as from \Eq{baromega}
\begin{eqnarray}
\label{VL}
\bar{\omega}_k^2&=& \omega_k^2+ \int_0^{\omega_P} d\omega \frac{|V_k(\omega)|^2 \bar{\omega}_k}{\omega}=\omega_k^2 \frac{q_P + \omega_{P}}{q_P},\\
\bar{\omega}_x^2&=& \tilde{\omega}_x^2+ \int_0^{\omega_M} d\omega \frac{|Q(\omega)|^2\bar{\omega}_x}{\omega}=\tilde{\omega}_x^2 \frac{q_M + \omega_{M}}{q_M}. \nonumber
\end{eqnarray}
The form of the couplings in \Eq{V} has been chosen to recover, in the limit of diverging cutoff frequencies, Lorentzian resonances centered at the bare excitation frequencies. This is shown in more details in Appendix \ref{appendix:AppB}, but by inserting \Eq{V} in \Eq{ze} we  obtain
\begin{eqnarray}
\lim_{\omega_P\rightarrow \infty} \zeta_k(\omega)&=& i\sqrt{\frac{2 \gamma_P \omega^3}{\pi}} \frac{1}{\omega^2-\omega_k^2-i\gamma_P\omega}, \\
\lim_{\omega_M\rightarrow \infty}\eta(\omega)&=& \sqrt{\frac{2 \gamma_M \omega}{\pi}} \frac{1}{\omega^2-\tilde{\omega}_x^2-i\gamma_M\omega},\nonumber
\end{eqnarray}
from which we recognise the two Lorentzian lineforms which would be obtained from a classical Lorentz dielectric model with center frequency $\omega_k$ or $\tilde{\omega}_x$ and width $\gamma_P$ or $\gamma_M$. Notice that the resonances are centered at the bare frequencies, not the ones renormalised from the interaction with the reservoirs in \Eq{VL}, which instead diverge with the cutoff frequency. 
Hence, according to \Eq{ZZZ} we can calculate the other two required functions
\begin{eqnarray}
Z(\omega)&=& 2  \frac{\omega^2-\tilde{\omega}_x^2}{\left(\omega^2-\tilde{\omega}_x^2\right)^2+\gamma_M^2 \omega^2},\\
W_k(\omega)&=& 2 \frac{\omega_k^2 \left(\omega^2-\omega_k^2\right)-\gamma_P^2 \omega^2}{\left(\omega^2-\omega_k^2\right)^2+\gamma_P^2 \omega^2}. \nonumber
\end{eqnarray}

\begin{figure}[ht!]
    \centering
    \includegraphics[width=0.5\textwidth]{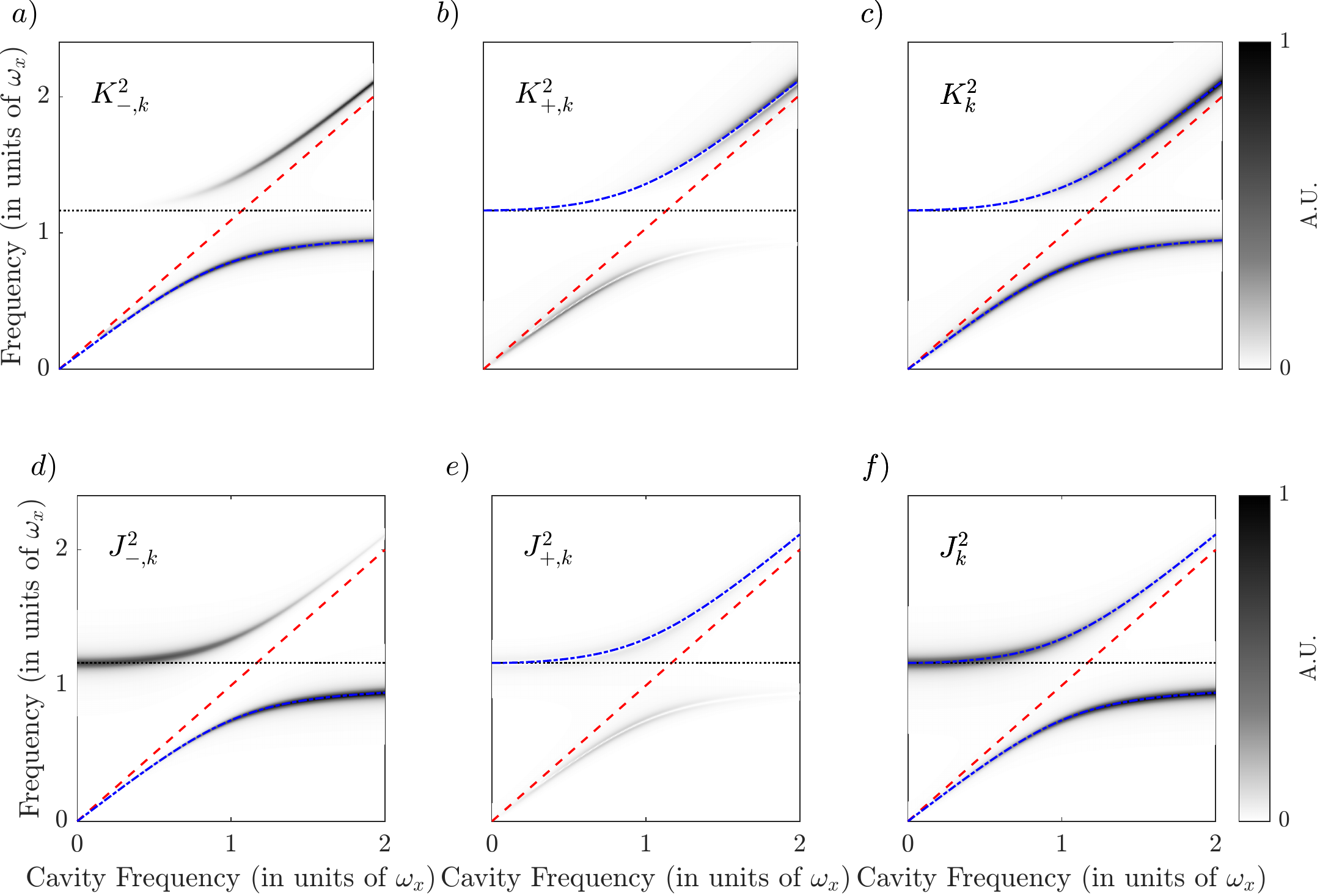}
    \caption{This figure displays the light $|K_{\pm,k}|^2$ (a)-(b) and matter $|J_{\pm,k}|^2$ (d)-(e) components of the two polaritonic branches, as wells as their sums $|K_{k}|^2$ (c), and $|J_{k}|^2$ (f), when the gauge is fixed by $s_{x,k,-}=0$.
    The field spectra are plotted as functions of the bare cavity frequency $\omega_k$ (red dashed line), while the resonant matter frequency $\tilde{\omega}_x$ is fixed and used as unit of frequency (black dotted line). Other parameters are: $g=0.3\omega_x$, $\gamma_P=\gamma_M=0.05\omega_x$. The field spectra are here normalised to the maximum value for all the plots in the same row.
    The calculated polariton modes in the lossless case $\omega_{-,k}$ and $\omega_{+,k}$ are marked by a dot-dashed blue lines.
    Although we maintain the notation $K_{\pm,k}$ and $J_{\pm,k}$, it is clear that it is no longer possible to isolate the contributions of the two polariton branches}
    \label{fig:K}
\end{figure}

\begin{figure}[ht!]
    \centering
    \includegraphics[width=0.48\textwidth]{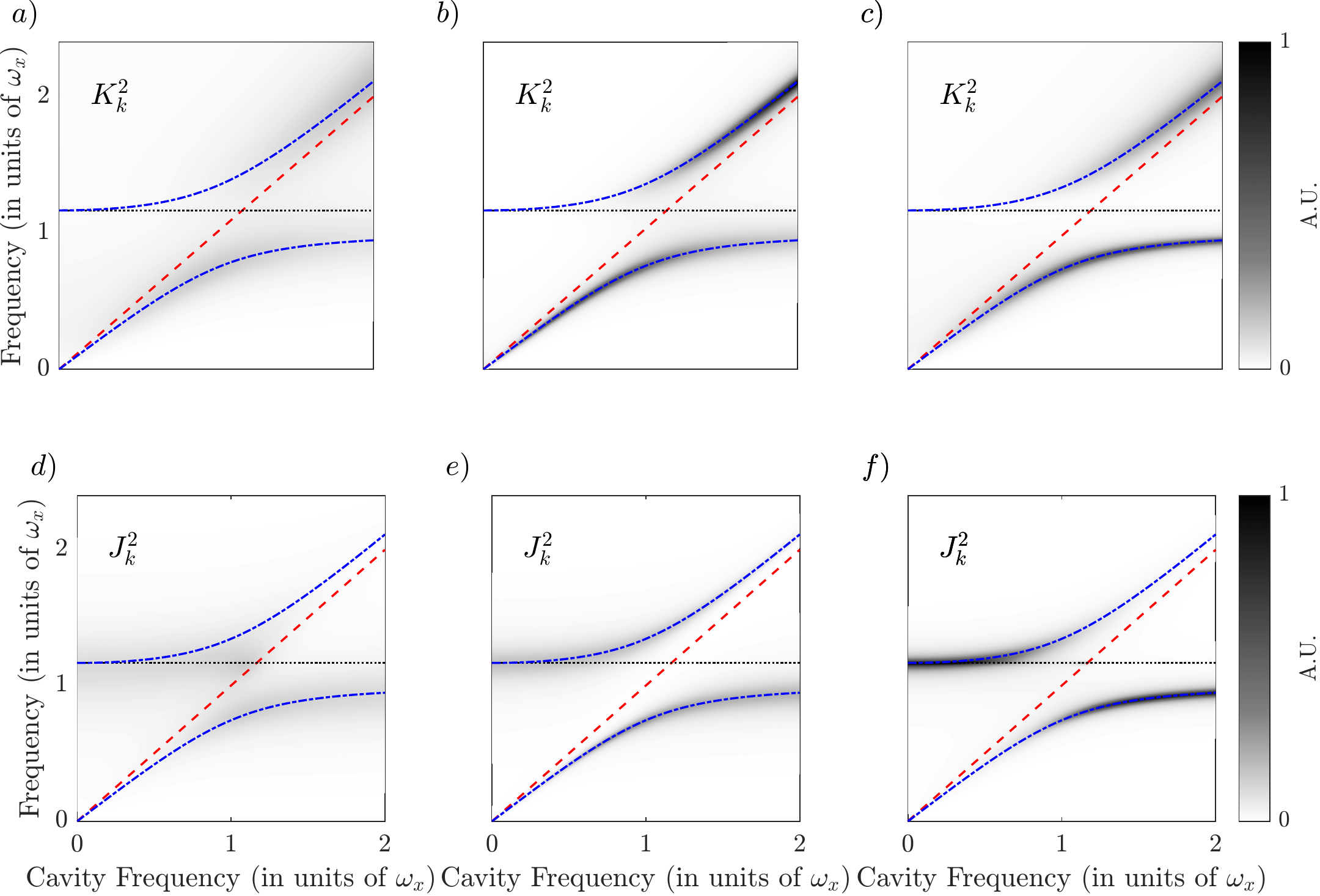}
    \caption{The panels display the effects of the interplay between light and matter losses  on the analytically derived dressed photonic  $|K_{k}(\omega)|^2$ (a)-(c), and matter $|J_{k}(\omega)|^2$ (b)-(d) fields. In all the plot $g=0.3\omega_x$, while the losses rates are: $\gamma_M=\gamma_P=0.4\omega_x$ in (a) and (d); $\gamma_P=0.05\omega_x$ and $\gamma_M=0.2\omega_x$ in (b) and (e); $\gamma_P=0.2$ and $\gamma_M=0.05\omega_x$ in (c) and (f). The field spectra are normalised to the maximum value for all the plots in the same row. All the other parameters remain as in Fig. \ref{fig:K}.}
    \label{fig:KJ}
\end{figure}

\begin{figure}[ht!]
    \centering
    \includegraphics[width=0.5\textwidth]{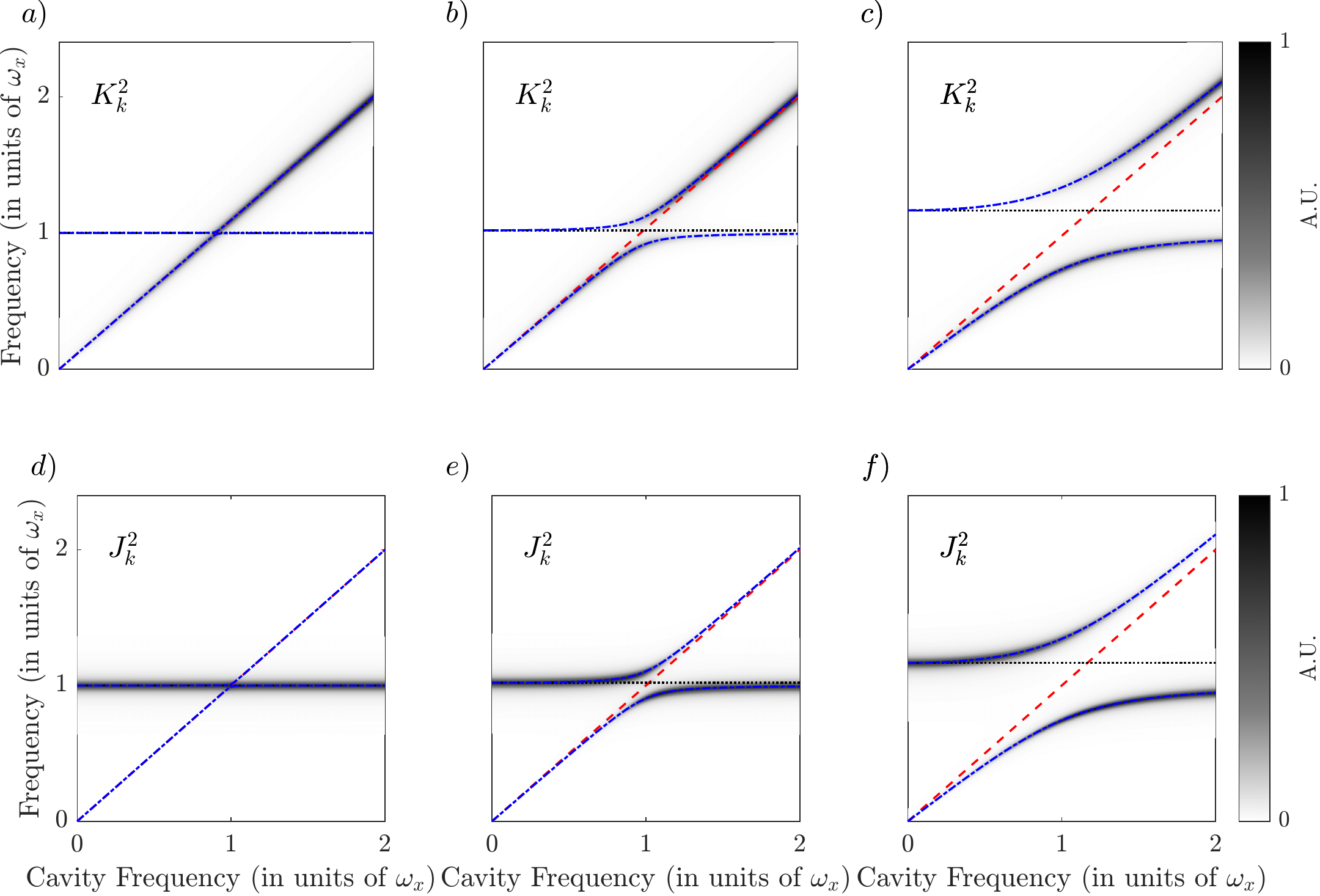}
    \caption{The panels display the analytical dressed photonic $|K_{k}(\omega)|^2$ (a)-(c) and matter $|J_{k}(\omega)|^2$ (b)-(d) fields varying the light-matter coupling strength $g$. In all the plot $\gamma_M=\gamma_P=0.05\omega_x$ , while the coupling is: $g=0.01\omega_x$ in (a) and (d); $g=0.1\omega_x$ in (b) and (e); $g=0.3\omega_x$ in (c) and (f). The field spectra are here normalised to the maximum value for all the plots in the same row. All the other parameters remain as in Fig. \ref{fig:K}.}
    \label{fig:KJG}
\end{figure}

Using these explicit forms for the couplings in Fig. \ref{fig:K} we plot the photonic and material component of each polaritonic branch obtained with the gauge-fixing $s_{x,k,-}(\omega)=0$ used in Ref. \cite{de_liberato_virtual_2017}.
As we can see the distinction between the operators of the two polaritonic branches, represented in the first two columns, is arbitrary and we cannot identify them with specific resonances. Only their sum, the total intensity of the photonic or matter components, shown in the third column, has physical meaning. 
 In all the figures the black dotted line represents the resonant frequency of the material resonance $\tilde{\omega}_x$, the red dashed line the photonic frequency $\omega_k$, and the black dash-dotted lines the polaritonic resonances in the lossless case from \Eq{lossless}.
 In Fig. \ref{fig:KJ} and Fig. \ref{fig:KJG} we plot the gauge-independent results from the \Eq{final} for different values of, respectively, the losses and the coupling strength, showing that the theory behaves as predicted from input-output theories \cite{savasta_quantum_1996,ciuti_input-output_2006,de_liberato_comment_2014}, with two polaritonic Lorentzian resonances with linewdiths proportional to the weighted average of the light and matter respective linewidths.     
 
\section{Diagonalization with coloured reservoirs}	
\label{Intrinsec}

\begin{figure}[ht!]
    \centering
    \includegraphics[width=0.48\textwidth]{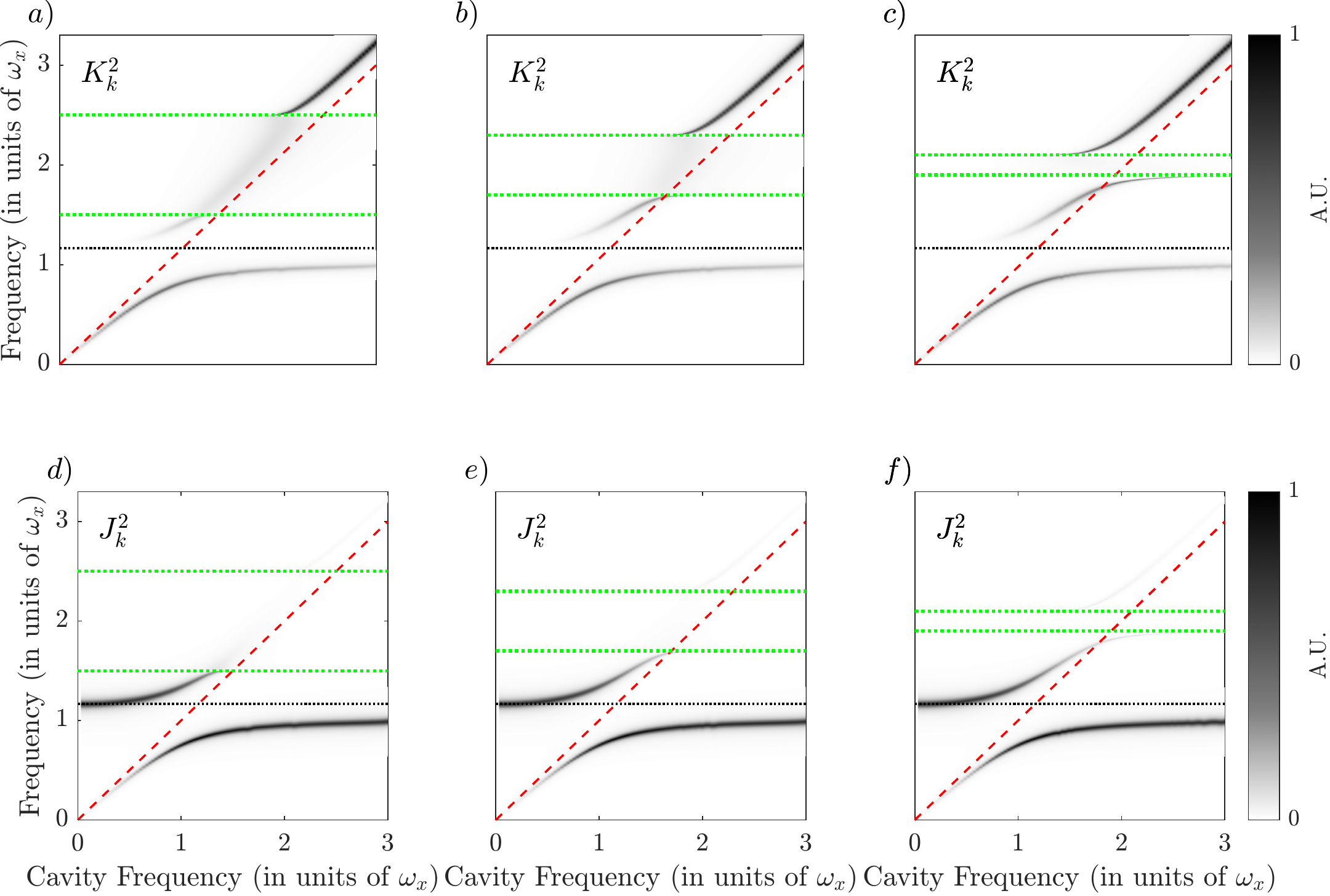}
    {\caption{The panels display the field functions $|K_{k}(\omega)|^2$ (a)-(c) and $|J_{k}(\omega)|^2$ (b)-(d), with the presence of a rectangular absorption band centered at frequency $\omega_c=2$ and of width $\Delta$. The green dotted lines mark the band boundaries. In all the plot $\kappa=0.05 \omega_x$, the reservoir losses rates are $\gamma_M=\gamma_P=0.05\omega_x$, and $\Delta=\omega_x$ (a) and (d), $\Delta=0.6\omega_x$ (b) and (e) and $\Delta=0.2\omega_x$ (c) and (f). The field spectra are normalised to the maximum value for all the plots in the same row.
    All the other features remain as in Fig. \ref{fig:K}.}}
    \label{fig:KJC}
\end{figure}

The theory we developed allows to model polaritonic systems with arbitrary coloured reservoirs, including the scientifically and technologically relevant case of a continuum absorption band, a case object of multiple theoretical \cite{averkiev_light-matter_2007,citrin_microcavity_2003,cortese_strong_2019,parish_excitons_2021,cao_strong_2021} and experimental \cite{liu_quantum_2017,mueller_deep_2020,cortese_excitons_2021,rajabali_polaritonic_2021} studies. 
Focusing for the sake of definiteness on a compactly-supported reservoir interacting with the photonic component of the excitation (an absorption band), we can include it in the theory by choosing frequency-dependent coupling functions $V_k(\omega)$ with support in the chosen frequency range. Exactly the same procedure would  couple to the matter component by using $Q(\omega)$ instead.

The presence of an absorption band will generally not substitute other loss channels influencing the excitations lifetime and the necessity of keeping both can cause some formal problem given that, as we saw before, the modelling of a Lorentzian resonance requires us to renormalize an otherwise infinite resonant frequency.
In this Section we will provide the recipe to add an arbitrary coloured reservoir on the top of the Lorentzian one.
The microscopic reservoir modes leading to the Lorentzian lineshape of the uncoupled photonic resonance are {\it a priori} completely uncorrelated from those leading to an absorption band. They will in many case have even different physical origins, {\it e.g.,} phonon interaction for the former and interaction with electronic bands for the latter. As better explained in Appendix \ref{appendix:Appz} this means their effects sum incoherently and we can thus write the full interaction of the photonic component with its environment using the coupling function
 \begin{equation} \label{VV}
 |V_k(\omega)|^2=|V^0_k(\omega)|^2+|V^1_k(\omega)|^2,
 \end{equation}
where $V^0_k(\omega)$  and $V^1_k(\omega)$ model, respectively,  the interaction with the photonic reservoir and the absorption band, defined by a normalised density $F(\omega)$ with $F(\omega<0)=0$, $\int_0^{\infty}d\omega F(\omega)=1$, and a dimensionless coupling intensity $\kappa$.
The two coupling functions take the form 
\begin{eqnarray} \label{V1}
|V^0_k(\omega)|^2&=&\frac{\omega {{\bar{\omega}_k}}}{q_P+\omega_{P}}\Theta(\omega_P-\omega),\\\nonumber
|V^1_k(\omega)|^2&=& \frac{\omega{{\bar{\omega}_k}}q_P }{q_P+ \omega_P} \frac{\kappa }{1+\kappa} F(\omega),
\end{eqnarray}
where $|V^0_k(\omega)|^2$, which will lead to the Lorentzian broadening as in the previous case, has the same form as in \Eq{V}, and the other term is chosen in order to provide the required absorption band after the renormalization.

The dressed frequency of the photonic mode is now renormalized by both terms.
It is practical to write the impact of the Lorentzian broadening as a renormalization over the frequency of the photonic resonance $\tilde{\omega}_k$ dressed only by the absorption band
\begin{eqnarray} \label{textwk}
{{\bar{\omega}_k^2}}&=& \tilde{\omega}_k^2+\int_0^\infty d\omega \frac{|V^0_k(\omega)|^2 {{\bar{\omega}_k}}}{\omega}=\tilde{\omega}_k^2 \frac{q_P+ \omega_P}{q_P},
\end{eqnarray}
and define the latter as 
\begin{eqnarray} \label{tilde}
\tilde{\omega}_k^2&=& \omega_k^2+\int_0^\infty d\omega \frac{|V^1_k(\omega)|^2 {{\bar{\omega}_k}}}{\omega}= \omega_k^2(1+\kappa).
\end{eqnarray}

We then follow the same diagonalization procedure as in Section \ref{reservoirs}, with function $\chi_k(\omega)$ now taking the form \begin{eqnarray}
 \chi_k(\omega)&=&1-\frac{1}{2 {{\bar{\omega}_k}}} \int_{-\infty}^\infty d\omega' \frac{\mathcal{V}^0_k(\omega')}{\omega'-\omega+i0^+}\\ \nonumber && -\frac{1}{2 {{\bar{\omega}_k}}} \int_{-\infty}^\infty d\omega' \frac{\mathcal{V}^1_k(\omega')}{\omega'-\omega+i0^+},\nonumber
\end{eqnarray}
where $\mathcal{V}^0_k(\omega)$ and $\mathcal{V}^1_k(\omega)$ are the odd analytical extensions in the negative frequency range of, respectively, $|V^0_k(\omega)|^2$ and $|V^1_k(\omega)|^2$.
The squared function $|\zeta_k(\omega)|^2$ can be written as in \Eq{ze}
\begin{eqnarray} \label{zk}
|\zeta_k(\omega)|^2&=& \frac{|V^0_k(\omega)|^2 \omega^2 {{\bar{\omega}_k}} }{|\omega^2-{{\bar{\omega}_k}}^2 \chi_k(\omega) |^2} + \frac{|V^1_k(\omega)|^2 \omega^2 {{\bar{\omega}_k}} }{|\omega^2-{{\bar{\omega}_k}}^2 \chi_k(\omega) |^2},\quad
\end{eqnarray}
leading to, after some algebra, and after letting $\omega_P\rightarrow \infty$, to
\begin{eqnarray} \label{Gamma}
|\zeta_k(\omega)|^2&=&  \frac{\omega^3}{\left[\omega^2-\Omega_k^2(\omega)\right]^2+\omega^2 \Gamma(\omega)^2}\left[\frac{2 \gamma_P}{\pi}+\kappa\omega_k^2 F(\omega)\right]\nonumber\\
\end{eqnarray}
with effective central frequency and effective losses
\begin{eqnarray}\label{OG}
&&\Omega_k^2(\omega)= \omega_k^2\left[1+\kappa\left(1-\frac{1}{2}P\int_{-\infty}^{\infty}\frac{\omega' F(|\omega'|)}{\omega'-\omega}  d\omega'
\right) \right],
 \nonumber\\
&&\Gamma(\omega)=\gamma_P+\frac{\pi\kappa\omega_k^2  F(\omega)}{2},
\end{eqnarray}
where the linewidth of the Lorentzian losses is now defined as a function of the frequency dressed by the absorption band $\gamma_P=\frac{\pi \tilde{\omega}_k^2}{2 q_P}$.
Note that in the low-frequency regime $\omega\rightarrow 0$ from \Eq{OG} we have $\Omega_k\approx \omega_k$, showing that the presence of the absorption band does not change the background permittivity.

From \Eq{Gamma} we see that the photonic losses increase in the frequency region in which $F(\omega)\neq 0$, an effect already observed in Ref. \cite{rajabali_polaritonic_2021}, and it also leads to a resonance effect in the central frequency. This can be understood in the light of recent works on strong coupling with the continuum  \cite{cortese_strong_2019,cao_strong_2021}, and we expect it to model the possibility of the photonic resonance becoming strongly coupled with the absorption band.
From the renormalised expressions in \Eq{OG}, the integral function $W_k(\omega)$ in \Eq{ZZZ} can be simply evaluated numerically.

In Fig. \ref{fig:KJC} we plot example results obtained using a sharp absorption band of center frequency $\omega_c$ and width $\Delta$
\begin{align}
F(\omega)&=\frac{1}{\Delta}\Theta \left(\frac{\Delta}{2}-|\omega-\omega_c|\right).
\end{align}
The boundaries of the band are marked by horizontal green dashed lines. We recognise the effects expected from our analytical results: the band acts as a localised absorber for the photonic component of the polariton and eventually the polaritonic mode gets strongly coupled to the band, an effect better visible when the band width becomes comparable with the intrinsic photonic linewidth.


\section{Conclusions}
In this article we exactly solved the polaritonic problem with a quantum formalism in the case of arbitrary dissipative couplings for both the bare photonic and matter excitations. In order to do this we discussed the extension of the Fano and HB theories to the case of multiple discrete levels coupled to multiple continua, showing how a gauge indeterminacy emerges. While a previous approach to this problem had been to peform an arbitrary gauge choice, here we analytically calculated the gauge-invariant observables. We thus demonstrated both the self-consistency of our theory and provided an analytical, albeit cumbersome formula allowing to exactly calculate the resonance lineshape for strongly coupled resonances interacting with reservoirs of arbitrary spectral shapes. We then showed how coloured features can pratically be added to an homogeneous resonance linewidth. Note that while our approach is based on a purely bosonic spectrum of the material resonance, generally correct for plasmonic and phononic systems, saturation and finite size effects can {\it a priori} be included in the theory as higher order terms \cite{cortese_polariton_2017}.

We hope these results will be of use in the subfields of polaritonic in which losses have an important impact. This is true for example in plasmonic systems, characterised by important Ohmic losses, nonlocal systems where photonic excitations couple to a continuum of propagative modes, and in systems in which the extremely large coupling between light and matter pushes the polaritonic resonances into other spectral features.

\section{Acknowledgements}
 S.D.L. is a Royal Society Research Fellow and was partly funded by the Philip Leverhulme Prize of the Leverhulme Trust. The authors acknowledge funding from the RGF81001 grant from the Royal Society.
 
\appendix

 \section{The Fano diagonalization and its extension} \label{appendix:Appz}
 In this Appendix we briefly discuss the basic idea behind the Fano diagonalization and its extenstions to the case of many cantinua or many discrete levels. 
Following Ref. \cite{fano_effects_1961} we consider the Hamiltonian
\begin{align}
\label{HF1}
H=\,& \omega a^{\dagger}a+\int\!d\omega'\, \omega' b^{\dagger}(\omega')b(\omega')\\&+
 \int\!d\omega'\, 
g(\omega')\left[b^{\dagger}(\omega')a+a^{\dagger}b(\omega')\right],\nonumber
\end{align}
where in the original paper  $a$ and $b(\omega)$ in \Eq{HF1} are normalised Hilbert space vectors, but in the present context they can as well be interpreted as second-quantized annihilation operators.
Under the assumption that all the coupled eigenvalues fall inside the initial continuum range Fano showed how the system can be exactly diagonalised in term of an hybridised continuum
\begin{align}
\label{PP}
P(\omega)=\,&x(\omega)a+\int\!d\omega' y(\omega,\omega')b(\omega').    
\end{align}
The discrete mode thus gets dressed by a cloud of continuum excitations, translating into a spectral broadening of the resonance.
Notice that this set of solution is not necessarily complete, as known from the study of the Friederichs-Lee model \cite{facchi_spectral_2021}.
In the bound-to-continuum strong coupling regime discrete modes can emerge from the continuum, as theoretically and experimentally demonstrated in the case of two-dimensional electron gases \cite{cortese_strong_2019,cortese_excitons_2021}.

After having completed the diagonalization procedure, Fano passes to consider the case in which there are $N$ discrete levels and one continuum. Such a problem can be reduced to the one treated above by  initially performing a partial diagonalization of one discrete level coupled to the continuum, leading to a novel Hamiltonian in the same form as the initial one but this time with $N-1$ discrete levels.
Proceding by iteration the system can be solved in term of a single hybridised continuum of the form 
\begin{align}
\label{P2}
P(\omega)=&\sum_{n=1}^N x_n(\omega)a_n+\int\!d\omega' y(\omega,\omega')b(\omega').    
\end{align}

Finally, the case of a single discrete state coupled to $N$ continua is treated, described by the Hamiltonian
\begin{align}
\label{H1N}
H=\,&\omega a^{\dagger}a+\sum_{n=1}^N\int\!d\omega'\, \omega' b_n^{\dagger}(\omega')b_n(\omega')\\&+
 \sum_{n=1}^N\int\!d\omega'\, 
g_n(\omega')\left[b_n^{\dagger}(\omega')a+a^{\dagger}b_n(\omega')\right].\nonumber
\end{align}
Such a system can be solved by
performing the transformation
\begin{align}
\label{tildeb}
\tilde{b}_m(\omega)&=\sum_{n=1}^N U_{mn}(\omega)b_n(\omega), \end{align}
where $U_{mn}(\omega)$ is a unitary matrix whose first row is given by
\begin{align} \label{Eq6}
U_{1n}(\omega)&=\,\frac{g_n(\omega)}{\tilde{g}_1(\omega)},
\end{align}
with
\begin{align}
\tilde{g}_1(\omega)&=\sqrt{\sum_{n=1}^{N}  g_n(\omega)^2},
\end{align}
transforming \Eq{H1N} into 
 the Hamiltonian of one discrete level coupled to a single continuum, plus other $N-1$ uncoupled continua 
\begin{align}
H=\,&\omega a^{\dagger}a+\sum_{n=1}^N\int\!d\omega'\, \omega' \tilde{b}_n^{\dagger}(\omega')\tilde{b}_n(\omega')\\&+
\int\!d\omega'\, 
\tilde{g}_1(\omega')\left[\tilde{b}_1^{\dagger}(\omega')a+a^{\dagger}\tilde{b}_1(\omega')\right].\nonumber
\end{align}

 \section{Huttner-Barnett  Diagonalization}
 \label{appendix:AppF}
In this Appendix we perform the HB diagonalization of the Hamiltonian $H_{\text{PB}}$ in \Eq{PB}, describing the interaction between the discrete photon mode and the photonic reservoir, modeled as an ensamble of harmonic oscillators indexed by the continuum frequency $\omega$.

We introduce the bosonic operators describing broadened photons $A_k(\omega) $, 
\begin{eqnarray} \label{A1}
A_k(\omega)&=& x_k(\omega) a_k+ z_k(\omega) a_k^\dagger + \\&&\int\!d\omega' \left [ y_k(\omega,\omega') \alpha_k(\omega')+ w_k(\omega,\omega') \alpha_k^\dagger (\omega') \right ], \nonumber 
\end{eqnarray}
whose coefficients are chosen so that the operators satisfy the eigenequation  
\begin{eqnarray}
\omega A_k(\omega)=\left[A_k(\omega),H_{\textrm{PB}}\right].
\end{eqnarray}
This equation leads to the system between the coefficients
\begin{eqnarray}
x_k(\omega) \left(\omega\! -\! \bar{\omega}_k \right )&=&\frac{1}{2}\int_0^\infty d\omega' \left [ y_k(\omega,\omega') V_k(\omega')- \right. \nonumber\\ && \left.w_k(\omega,\omega') V_k^*(\omega') \right],\label{xk} \\
z_k(\omega) \left(\omega\! +\! \bar{\omega}_k\right )&=&\frac{1}{2}\int_0^\infty d\omega' \left [ y_k(\omega,\omega') V_k(\omega')- \right. \nonumber\\ && \left. w_k(\omega,\omega') V_k^*(\omega') \right],\label{zzk}\\
y_k(\omega,\omega') \left(\omega\!-\!\omega'\right)&=&\frac{1}{2}\left [ x_k(\omega) - z_k(\omega)\right] V_k^*(\omega'), \label{Yk}\\
w_k(\omega,\omega') \left(\omega\!+\!\omega'\right)&=&\frac{1}{2}\left [ x_k(\omega) - z_k(\omega)\right] V_k(\omega').
\label{wk}
\end{eqnarray}
This set of equation can be solved to obtain $z_k(\omega)$, $y_k(\omega,\omega')$ and $z_k(\omega,\omega')$ in terms of $x_k(\omega)$. By subtracting \Eq{zzk} from \Eq{xk}, we obtain
\begin{eqnarray}
z_k(\omega)&=& \frac{\omega-\bar{\omega}_k}{\omega+\bar{\omega}_k} x_k(\omega),
\end{eqnarray}
which can be substituted in \Eq{Yk} and \Eq{wk} to obtain
\begin{eqnarray}\label{yk1}
y_k(\omega,\omega')&=& \left [P \left(\frac{1}{\omega-\omega'}\right) +\gamma_k(\omega) \delta(\omega-\omega')  \right] \times \\
&& V_k^*(\omega')\frac{\bar{\omega}_k}{\omega+\bar{\omega}_k} x_k(\omega),\nonumber\\ \nonumber
w_k(\omega,\omega')&=& \left [\frac{1}{\omega+\omega'}  \right]  V_k(\omega')\frac{\bar{\omega}_k}{\omega+\bar{\omega}_k} x_k(\omega). 
\end{eqnarray}

The function $\gamma_k(\omega)$ can be found, after some algebra, replacing both the equations in \Eq{yk1}  in \Eq{xk}
\begin{eqnarray} \label{gammak}
\gamma_k(\omega)\!&=&\!\frac{2(\omega^2\!-\!\bar{\omega}_k^2)}{\bar{\omega}_k|V_k(\omega)|^2} \!+ \!\frac{1}{|V_k(\omega)|^2} P\!\int_{-\infty}^\infty d\omega'\frac{ \mathcal{V}_k(\omega')}{\omega'-\omega},\quad
\end{eqnarray}
where we assume that the analytic extension in the negative frequency range $\mathcal{V}_k(\omega)$ of $|V_k(\omega)|^2$ is an odd function. In order to calculate $x_k(\omega)$ we impose the commutation relation
\begin{eqnarray} \label{Acomm}
\left[A_k(\omega),A_{k'}(\omega')^{\dagger} \right]&=&\delta_{k,k'}\delta(\omega\!-\!\omega').
\end{eqnarray}
By using the expression for $A_k(\omega)$ in \Eq{A1} and for the coefficients in \Eq{zzk} and \Eq{yk1}, in terms of $x_k(\omega)$, \Eq{Acomm} leads to definition of $x_k(\omega)$ up to a phase factor 
\begin{eqnarray} \label{xk11}
x_k(\omega)&=& \frac{\omega+\bar{\omega}_k}{\bar{\omega} V_k^*(\omega)}\frac{1}{\gamma_k(\omega)-i\pi}.
\end{eqnarray}
Exploiting the expression for $\gamma_k(\omega)$ in \Eq{gammak}, \Eq{xk11} can be written as 
\begin{eqnarray} 
x_k(\omega)&=&\frac{  \omega+\bar{\omega}_k }{2}  \frac{V_k(\omega)}{\omega^2-\bar{\omega}_k^2 \chi_k(\omega)},
\end{eqnarray}
with 
\begin{eqnarray} \label{chik1}
\chi_k(\omega)=1-\frac{1}{2 \bar{\omega}_k} \int_{-\infty}^\infty d\omega' \frac{\mathcal{V}_k(\omega')}{\omega'-\omega+i 0^+}.
\end{eqnarray}

The final expressions for the coefficients are thus obtained as 
\begin{eqnarray} \label{xx}
 x_k(\omega)&=&\frac{  \omega+\bar{\omega}_k }{2}  \frac{V_k(\omega)}{\omega^2-\bar{\omega}_k^2 \chi_k(\omega)},\\ \nonumber
 z_k(\omega)&=&\frac{  \omega-\bar{\omega}_k }{2}  \frac{V_k(\omega)}{\omega^2-\bar{\omega}_k^2 \chi_k(\omega)},\\ \nonumber
 y_k(\omega,\omega')&=& \delta(\omega\!-\!\omega')+\\&& \frac{\bar{\omega}_k}{2} \frac{V_k(\omega')}{\omega\!-\!\omega'-i 0^+} \frac{V_k(\omega)}{\omega^2-\bar{\omega}_k^2 \chi_k(\omega)},\nonumber\\ \nonumber
 w_k(\omega,\omega')&=&  \frac{\bar{\omega}_k}{2}  \frac{V_k(\omega')}{\omega\!+\!\omega'} \frac{V_k(\omega)}{\omega^2-\bar{\omega}_k^2 \chi_k(\omega)}. \end{eqnarray}

 \section{Diagonalization in the Coulomb representation} \label{appendix:AppA}
 \begin{figure}[ht!]
    \centering
    \includegraphics[width=0.5\textwidth]{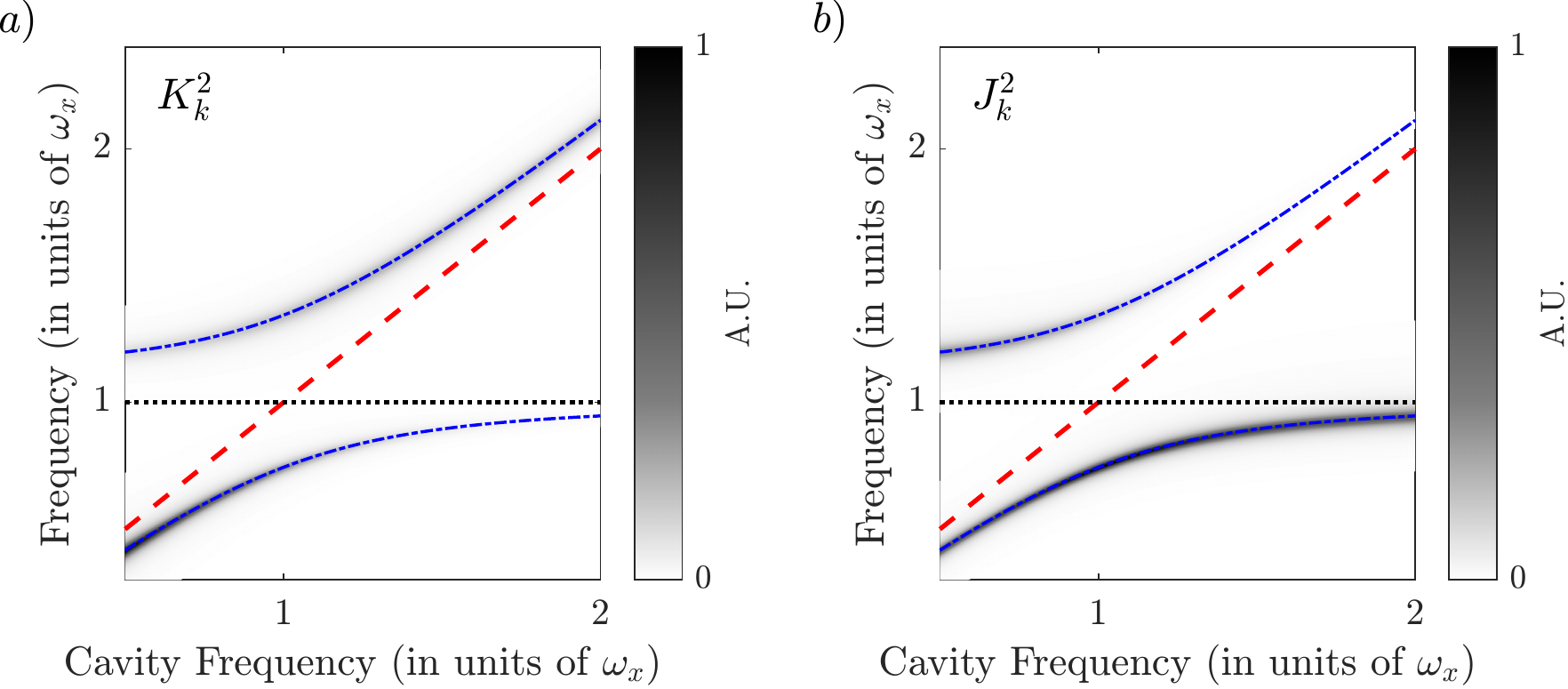}
    \caption{Coupled photonic $K_{k}^2$ (a) and matter $J_{k}^2$ (b) fields as a function of the bare cavity frequency $\omega_k$, calculated starting from a Coulomb representation Hamiltonian. The light-matter couling is $g=0.3\omega_x$, while the loss rates are $\gamma_P=\gamma_M=0.05\omega_x$. All the other parameters remain as in Fig. \ref{fig:K}.}
    \label{fig:KJCou}
\end{figure}
In this Appendix, we show how our approach gets modified if one wants to work in the Coulomb representation. Given that only relatively few quantities are affected by the change, we just provide the expressions for the affected ones.
The Hamiltonian  in the Coulomb representation can be written as
\begin{eqnarray}
		H&=& \sum_k \left (\omega_k a^{\dagger}_k a+\omega_x \, b^{\dagger}_k b_k \right )
		+ \\&&i \sum_k g \sqrt{\frac{\omega_x}{\omega_k}}\left[a^{\dagger}_k+a_k \right]
		\left[b^{\dagger}_k-b_k\right]
		+
		\sum_k \frac{\lvert g\rvert^2}{\omega_k}
		\left[a^{\dagger}_k+a_k \right]^2, \nonumber \label{col}
	\end{eqnarray}
in which we can see the appearance of a diamagnetic $A^2$ term.
After reabsorbing such a diamagnetic term by performing a Bogoliubov transformation, the Hamiltonian takes the form
\begin{eqnarray}\label{Ha}
		H&=& \sum_{k} \left (\tilde{\omega}_k a^{\dagger}_k a+\omega_x \,b^{\dagger}_k b_k \right)
		+\\&& i \sum_{k} \sqrt{\frac{\omega_x}{\bar{\omega}_k}} \left[a^{\dagger}_k+a_k \right]
		\left[g \, b_{k}^{\dagger}-g^* \, b_k\right], \nonumber
	\end{eqnarray}
with renormalised cavity frequency 

\begin{eqnarray}
\tilde{\omega}_k^2=\omega_k^2+4|g_k|^2.
\end{eqnarray}
The dispersion relation obtained diagonalizing the Hamiltonian in \Eq{Ha} is
	\begin{equation}
\omega_{\pm,k}^2-\tilde{\omega}_k^2= \frac{4 |g|^2\omega_x^2}{\omega_{\pm,k}^2-\omega_x^2},
\end{equation}
with solutions
\begin{equation}
\omega_{\pm,k}= \frac{1}{\sqrt{2}}\sqrt{\tilde{\omega}_k^2+ \omega_x^2 \pm \sqrt{(\tilde{\omega}_k^2-\omega_x^2)^2 + 16 |g|^2\omega_x^2}}.
\end{equation}
In the Coulomb representation the expressions of the field operators
in \Eq{fieldsab} are modified as
\begin{eqnarray}
\left (a_k+a_k^\dagger \right)&=&\sqrt{\bar{\omega}_k} \int_0^\infty d\omega \left [ \zeta_k(\omega) A_k^\dagger(\omega) + \zeta_k^*(\omega) A_k(\omega) \right], \nonumber\\ 
i \left(b^\dagger_k- b_k \right)&=&\frac{1}{\sqrt{\bar{\omega}_x}}\int_0^\infty d\omega' \left [\eta(\omega) B_k^\dagger(\omega)+\eta^*(\omega) B_k(\omega) \right ],\nonumber \\ 
\end{eqnarray}
where
\begin{eqnarray} \label{zc}
\zeta_k(\omega)&=&\frac{1}{\sqrt{\bar{\omega}_k}} \left[x_k(\omega)-z_k(\omega)\right]=\frac{V_k(\omega) \sqrt{\bar{\omega}_k }}{\omega^2-\bar{\omega}_k^2 \chi_k(\omega)},\\ \nonumber
\eta(\omega)&=&i \sqrt{\bar{\omega}_x} \left [\bar{x}(\omega)+\bar{z}(\omega) \right]=i \frac{Q(\omega) \omega \sqrt{\bar{\omega}_x }}{\omega^2-\bar{\omega}_x^2t(\omega)}.
\end{eqnarray}
By substituting the bare operators with the dressed ones, we arrive to the Hamiltonian as in \Eq{HTOT}, which can be diagonalised by the same procedure described in the main text. 
The dressed photonic and matter field operator can finally be written as superpositions of polaritonic broadened modes as
\begin{eqnarray}
\left (a_k+ \!a_k^\dagger\right)&=& \!\sqrt{\bar{\omega}_k}\int^{\infty}_0  d\omega \! \sum_{j=\pm}\!\left[ K^*_{k,j}(\omega) P_j(\omega)\!+\right.
\\ && \left.  K_{k,j}(\omega) P_j^\dagger(\omega) \right],\nonumber\\ 
i\left(b_k-\! b_k^\dagger\right)&=& \! \frac{1}{\sqrt{\bar{\omega}_x}}\int^{\infty}_0  d\omega \!  \sum_{j=\pm}   \left[ J^*_{k,j}(\omega) P_j(\omega)\!\nonumber+\right.
\\ && \left.  \! J_{k,j}(\omega)  P_j^\dagger(\omega) \right],\nonumber 
\end{eqnarray}
where
\begin{align}
K_{k,j}(\omega)&=\frac{1}{\sqrt{\bar{\omega}_k}}\left[\mathsf{X}_{k,j}(\omega)-\mathsf{Z}_{k,j}(\omega)\right],\\ \nonumber J_{j}(\omega)&=-i {\sqrt{\bar{\omega}_x}} \left[\mathsf{Y}_{j}(\omega)+\mathsf{W}_{j}(\omega) \right].
\end{align}
Note that in the Coulomb representation the functions related to the photonic component and those related to the matter part have exchanged units from those in the PZW representation.  This is due to the inverted dependence of the light and matter fields upon their frequency.

\section{Derivation of analytical results for a Lorenztian broadening} \label{appendix:AppB}
In this Appendix we will derive the analytical expression of the functions $\zeta_k(\omega)$ and $\eta(\omega)$ in the case of a Lorentzian broadening.
This is not completely trivial, as testify by the existence of a published paper claiming it is impossible \cite{dutra_permittivity_1998}. The key issue is that in order to recover a frequency-independent broadening, the shift in the mode due to the coupling with the reservoir has to diverge. As such the result can only be found by a renormalization procedure allowing to cancel such a divergence.

Assuming that the coupling potentials to the photonic and matter reservoirs take the form in \Eq{V}, we can calculate the dressed resonance frequencies 
\begin{eqnarray}
\bar{\omega}_k^2&=& \omega_k^2+ \int_0^\infty d\omega \frac{|V_k(\omega)|^2 \bar{\omega}_k}{\omega}=\omega_k^2 \frac{q_P + \omega_{P}}{q_P},\\
\bar{\omega}_x^2&=& \tilde{\omega}_x^2+ \int_0^\infty d\omega \frac{|Q(\omega)|^2\bar{\omega}_x}{\omega}=\tilde{\omega}_x^2 \frac{q_M + \omega_{M}}{q_M},\nonumber
\end{eqnarray}
and the real and imaginary parts of functions $\chi_k(\omega)$ and $t(\omega)$ as 
\begin{eqnarray}
\text{Re}[\chi_k(\omega)]&=& \text{lim}_{\omega_{P}\rightarrow \infty}   \left [1- \frac{1}{2} \frac{1}{q_P+\omega_{P}} \times \right.  \\ \nonumber && \left.
\left (2 \omega_{P}+ 2 \omega \log \left ( 1-\frac{2 \omega}{\omega+\omega_{P}}\right) \right) \right],\\ \nonumber
\text{Im}[\chi_k(\omega)]&=& \text{lim}_{\omega_{P} \rightarrow \infty}  \frac{\pi \omega}{2 (q_P+\omega_{P})},\\ \nonumber
\text{Re}[t(\omega)]&=& \text{lim}_{\omega_M \rightarrow \infty}   \left [1- \frac{1}{2} \frac{1}{q_M+\omega_M} \times \right. \\ \nonumber &&\left. \left (2 \omega_M+ 2 \omega \log \left ( 1-\frac{2 \omega}{\omega+\omega_M}\right) \right) \right],\\\nonumber
\text{Im}[t(\omega)]&=& \text{lim}_{\omega_M \rightarrow \infty}  \frac{\pi \omega}{2 (q_M+\omega_M)}.
\end{eqnarray}
In the limit of infinite cut off frequency $\omega_{P} \rightarrow \infty$ and $\omega_M \rightarrow \infty$, the resonant frequencies diverge, but the intensity of the coupling vanishes and we arrive to the finite results 
\begin{eqnarray} \label{Rechi}
\text{lim}_{\omega_{P} \rightarrow \infty} \bar{\omega}_k^2 \text{Re}[\chi_k(\omega)]&=& \bar{\omega}_k^2 \left (\frac{q_P}{q_P+\omega_{P}} \right)= \omega_k^2,\\ 
\text{lim}_{\omega_{P} \rightarrow \infty} \bar{\omega}_k^2\text{Im}[\chi_k(\omega)]&=& \omega_k^2 \frac{\pi \omega}{2 q_P}.\nonumber\\ \nonumber 
\text{lim}_{\omega_M \rightarrow \infty} \bar{\omega}_x^2 \text{Re}[t(\omega)]&=& \tilde{\omega}_0^2 \left (\frac{q_M}{q_M+\omega_M} \right)= \tilde{\omega}_x^2,\\ \nonumber
\text{lim}_{\omega_M \rightarrow \infty} \bar{\omega}_x^2 \text{Im}[t(\omega)]&=& \tilde{\omega}_x^2\frac{\pi \omega}{2 q_M}.\nonumber
\end{eqnarray}
By inserting \Eq{Rechi} in \Eq{ze}, we finally obtain the Lorentzian form for the functions $\zeta_k(\omega)$ and $\eta(\omega)$ as 
\begin{eqnarray}
\zeta_k(\omega)&=& i\sqrt{\frac{2 \gamma_P \omega^3}{\pi}} \frac{1}{\omega^2-\omega_k^2-i\gamma_P\omega}, \\
\eta(\omega)&=& g\sqrt{\frac{2 \gamma_M \omega}{\pi}} \frac{1}{\omega^2-\tilde{\omega}_x^2-i\gamma_M\omega}.
\end{eqnarray}

\bibliography{main_PZW}
\end{document}